\documentclass{llncs}
\pdfoutput=1
\usepackage{makeidx}
\usepackage{graphicx}
\usepackage{url}
\usepackage{amssymb}

\newcommand{\pgm}[1]{{\small\texttt{#1}}}

\newcommand{\eof}{\ensuremath{\mathit{eof}}}
\newcommand{\below}{\sqsubseteq}

\newcommand{\tru}{\emph{true}}
\newcommand{\fls}{\emph{false}}
\newcommand{\ns}[1]{\mathit{ns}({#1})}
\newcommand{\qsh}{\ensuremath{q_\mathit{sh}}}
\newcommand{\qdh}{\ensuremath{q_\mathit{dh}}}

\begin{document}

\title{Static Analysis of File-Processing Programs using File Format Specifications}

\author{M. Raveendra Kumar\inst{1} \and Raghavan Komondoor\inst{2} \and
  S. Narendran\inst{2}}

\institute{
  Indian Institute of Science, Bangalore and Tata Consultancy
  Services Ltd.,
\email{raveendra.kumar@tcs.com}
\and
  Indian Institute of Science, Bangalore, 
  \email{$\{$raghavan,narendran$\}$@csa.iisc.ernet.in}
}

\maketitle

\begin{abstract}
Programs that process data that reside in files are widely used in varied
domains, such as banking, healthcare, and web-traffic analysis.  Precise
static analysis of these programs in the context of software verification
and transformation tasks is a challenging problem.  Our key insight
is that static analysis of file-processing programs can be made more useful
if knowledge of the input file formats of these programs is made available
to the analysis.  We propose a generic framework that is able to perform
any given underlying abstract interpretation on the program, while
restricting the attention of the analysis to program paths that are
potentially feasible when the program's input conforms to the given file
format specification.  We describe an implementation of our approach, and
present empirical results using real and realistic programs that show how
our approach enables novel verification and transformation tasks, and also
improves the precision of standard analysis problems.

\end{abstract}

\section{Introduction}
\label{sec:introduction}
Processing data that resides in files or documents is a central aspect of computing in many organizations and enterprises. Standard \emph{file   formats} or \emph{document formats} have been developed or evolved in various domains to facilitate storage and interchange of data, e.g., in banking~\cite{CLIEOP03,DTASPECS}, health-care~\cite{HL7STANDARDS}, enterprise-resource planning (ERP)~\cite{UNEDIFACT}, billing~\cite{fisherPads2011}, and web-traffic analysis~\cite{CLF}. The wide adoption of such standard formats has led to extensive development of software that reads, processes, and writes data in these formats.
However, there is a lack of tool support
for developers working in these domains that specifically targets the idioms commonly present in file-processing programs. We address this issue by proposing a generic approach for static analysis of file-processing programs that takes a program as well as a specification of the input file format of the program as input, and analyzes the program in the context of behaviors of the program that are compatible with the file-format specification.

\subsection{Motivating example}
\label{ssec:motivating-example}

\begin{figure}
\begin{minipage}{3.9in}
  \begin{tabular}{p{3.2in}p{0.7in}}
\setlength{\tabcolsep}{1pt}
\begin{tabular}{l}
\includegraphics[width=3.2in]{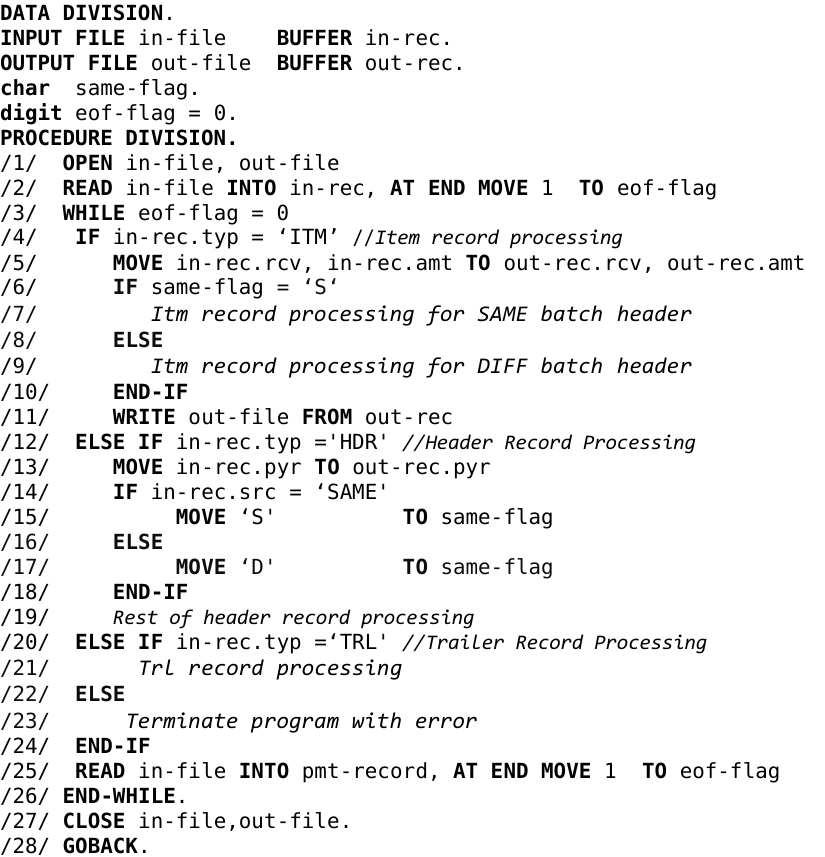}\\
\multicolumn{1}{c}{(a)}
\end{tabular}
&
\multicolumn{1}{l}{\begin{minipage}{0.7in}
\begin{tabular}{c}
\begin{scriptsize}
\begin{tabular}{|c|r|r|r|}
\hline
HDR & 10205 & 9000 &\multicolumn{1}{c|}{SAME} \\ \hline
ITM & 10201 & 3000 &\\ \hline
ITM & 10103 & 4000 &\\ \hline
ITM & 18888 & 2000 &\\ \hline
TRL & \multicolumn{3}{c|}{}            \\ \hline

HDR & 20221 & 6000 &\multicolumn{1}{c|}{DIFF} \\ \hline
ITM & 19999 & 2000 &\\ \hline
ITM & 10234 & 4000 &\\ \hline
TRL & \multicolumn{3}{c|}{}            \\ \hline
\end{tabular} 

\end{scriptsize}
\\
\multicolumn{1}{c}{(b)}
\\
\begin{scriptsize}
\begin{tabular}{cccc}
\texttt{typ} & \texttt{pyr} & \texttt{tot} & \texttt{src} \\\hline
\multicolumn{1}{|c|}{\rule[0.8em]{0em}{0em}HDR} & \multicolumn{1}{c|}{10205} &
\multicolumn{1}{c|}{9000} & \multicolumn{1}{c|}{SAME} \\\hline\\

\texttt{typ} & \texttt{rcv} & \texttt{amt} & \\\cline{1-3}
\multicolumn{1}{|c|}{\rule[0.8em]{0em}{0em}ITM} & \multicolumn{1}{c|}{10201} &
\multicolumn{1}{c|}{3000} & \\\cline{1-3}\\
\end{tabular}
\end{scriptsize}
\\
\multicolumn{1}{l}{\centering
\texttt{typ}: \emph{main type}.
}
\\
\multicolumn{1}{l}{\centering
\texttt{pyr}: \emph{payer account number}.
}
\\
\multicolumn{1}{l}{\centering
\texttt{tot}: \emph{total batch amount}.
}
\\
\multicolumn{1}{l}{\centering
\texttt{src}: \emph{source bank}.
}
\\
\multicolumn{1}{l}{\centering
\texttt{rcv}: \emph{receiver account num}.
}
\\
\multicolumn{1}{l}{\centering
\texttt{amt}: \emph{item amount}.
}
\\
\multicolumn{1}{c}{(c)}
\end{tabular}
\end{minipage} 
}
\end{tabular} 

 \end{minipage}
  \caption{(a) Example program. (b) Sample input file. (c) Input file record layouts.}
 \label{fig:RunningPgm}
\end{figure}

Our work has been motivated in particular by \emph{batch programs} in the context of enterprise legacy systems. Such programs are typically executed periodically, and in each run process an input file that contains ``transaction'' records that have accumulated since the last run.
In order to motivate the challenges in analyzing file-processing programs,  we introduce as a running example a small batch program, as well as a sample file that it is meant to process, in Figure~\ref{fig:RunningPgm}.

\paragraph{Input file format.}

Although our example is a toy one, the sample file shown (in Figure~\ref{fig:RunningPgm}(b)) adheres to a  simplified version of a real banking format~\cite{DTASPECS}.
Each record is shown as a row, with 
fields being demarcated by vertical lines. In this file format, the records are grouped logically into ``batches'', with each batch representing a group of ``payments'' from one customer to other customers. Each batch consists of a ``header'' record (value `{HDR}' in the first field), which contains information about the paying customer, followed by one or more ``item'' or ``payment'' records (`{ITM}' in the first field), which identify the recipients, followed by a ``trailer'' record (`{TRL}').  Figure~\ref{fig:RunningPgm}(c) gives the names of the fields of header as well as item records.
Other than the first field \pgm{typ}, which we have discussed above, another field of particular relevance to our discussions is the \pgm{src} field in header records, which identifies whether the paying customer is a customer of the bank that's running the program (`SAME'), or of a different bank (`DIFF'). The meanings of the other fields are explained in part~(c) of the figure.

\paragraph{The code.}
Figure~\ref{fig:RunningPgm}(a) shows our example program, which is in a Cobol-like syntax. The ``\texttt{DATA DIVISION}'' contains the declarations of the variables used in the program, including the input file buffer \texttt{in-rec} and output file buffer \texttt{out-rec}. \texttt{in-rec} is basically an overlay (or \emph{union}, following the terminology of the C language), of the two record layouts shown in Figure~\ref{fig:RunningPgm}(c). After any record is read into this buffer the program interprets  its contents using the appropriate layout based on the value of the \texttt{typ} field. The output buffer \texttt{out-rec} is assumed to have fields \texttt{pyr}, \texttt{rcv}, \texttt{amt}, as well other fields that are not relevant to our discussions. These  field declarations have been elided in the figure for brevity.

\sloppypar The statements of the program appear within the ``\texttt{PROCEDURE DIVISION}''. The program has a
\emph{main loop}, in lines~3-26. A record is read from the input file first outside the main loop (line~2), and then once at the end of each iteration of the loop (line~25). In each iteration the most recent record that was read is processed according to whether it is a header record (lines~12-19), item record (lines~4-11), or trailer record (lines~20-21). The sole \pgm{WRITE} statement in the program is in the item-record processing block (line~11), and writes out a ``processed'' payment record using information in the current item record as well as in the previously seen header record.
Lines~7 and~9 represent code (details elided) that populates certain fields of \texttt{out-rec} in distinct ways depending on whether the paying customer is from the same bank or a different bank.

\subsection{Analysis issues and challenges}
\label{ssec:analysis-issues}

File-processing programs typically employ certain idioms that distinguish them from programs in other domains. These programs read an unbounded number of input records, rather than have a fixed input size. Furthermore, typically, a program is designed not to process arbitrary inputs, but only input files that adhere to a known (domain related) file-format. State variables are used in the program to keep track of the types  of the records read until the current point of execution. These state variables are used to decide how to process any new record that is read from the file. For instance, in the program in Figure~\ref{fig:RunningPgm}(a), the variable \pgm{same-flag} is set in lines~14-17 to `\pgm{S}' (for ``same'') or to `\pgm{D}' (for ``different''), based on \pgm{src} field of the header record that has just been read. This variable is then used in line~6 to decide how to process item records in the same batch that are subsequently read. In certain cases, the state variables could also be used to identify unexpected or ill-formed sequences,  and to ``reject'' them. 

Analyzing, understanding, and transforming file-processing programs in precise ways requires a unique form of \emph{path sensitive} analysis, in which, at each program point, distinct information about the program's state needs to be tracked corresponding to each distinct pattern of record types that could have been read so far before control reaches the point. We illustrate this using example questions about the program in Figure~\ref{fig:RunningPgm}, answers to which would enable various verification and transformation activities.

\paragraph{Does the program silently ``accept'' ill-formed inputs?}
This is a natural and important verification problem in the context of file-processing programs. In our running example, if an (ill-formed) input file happens to contain an item record as the first record (without a preceding header record), the variable \pgm{same-flag} would be uninitialized after this item record is read and  when control reaches line~6. This is because this variable is initialized only when a header record is seen, in lines~14-17. Therefore, the condition in line~6 could evaluate non-deterministically. Furthermore, the output buffer \pgm{out-rec}, which will be written out in line~11, could contain garbage in its \pgm{pyr} field, because this field also is initialized only when a header record is seen (in line~13).

In other words, file-processing programs could silently write out garbage values into output files or databases when given ill-formed inputs, which is undesirable. Ideally, in the running example, the programmer ought to have employed an additional state variable (e.g., \pgm{hdr-seen}), to keep track of whether a header record is seen before every item record, and ought to have emitted a warning or aborted the program upon identifying any violation of this requirement. In other words, state tracking in file-processing programs is complex, and prone to being done erroneously.
Therefore, there is a need for an automated analysis that can check whether a program ``over accepts'' bad files (i.e., files that don't adhere to a user-provided specification of well-formed files).

\paragraph{What program behaviors are possible with well-formed inputs?}
In other situations, we are interested only in information on program states that can arise after (prefixes of) well-formed files have been read. For instance, a developer might be interested in knowing about possible uses of unitialized variables during runs on well-formed files only, without the clutter caused by warning reports pertaining to runs on ill-formed files. Intuitively, only the first category of warnings mentioned above signifies genuine errors in the program. This is because in many cases developers do not try to ensure meaningful outputs for corrupted input files.  In our example program, there are in fact no instances of uninitialized variables being used during runs on well-formed files.

On a related note, one might want to know if a program can falsely issue an ``ill-formed input'' warning even when run on a well-formed file. This sort of ``under acceptance'' problem could happen either due to a programming error, or due to misunderstanding on the developer's part as to what inputs are to be expected. This could be checked by asking whether statements in the program that issue these warnings, such as line~23 in the example program, are reachable during runs on well-formed inputs. In the example program it turns out that this cannot happen.

\paragraph{What program behaviors are possible under restricted scenarios of interest?}
In some situations there is a need to identify paths in a program that are taken during runs on certain narrower sub-classes of well-formed files. For instance, in the running example, we might be interested only in the parts of the program that are required when input files contain batches whose header records always have `\pgm{SAME}' in the \pgm{src} field; these parts constitute all the lines in the program except lines~6,~9, and~14-17 (variable \pgm{same-flag} will no longer need to be set or used because all input batches are guaranteed to be `\pgm{SAME}' batches). This is essentially a classical \emph{program specialization} problem, but with a file-format-based specialization criterion (rather than a standard criterion on the parameters to the program). Program specialization has various applications~\cite{conselPEAppl1998}, for example, in program comprehension, decomposition of monolithic programs to collections of smaller programs that have internally cohesive functionality, and reducing run-time overhead.

\subsection{Our approach and contributions}
\label{ssec:intro:approach}

\paragraph{Approach: Static analysis based on file formats.}
The primary contribution of this paper is a generic approach to perform any given ``underlying'' abstract interpretation~\cite{Cousot77} of interest $U$, based on an abstract lattice $L$, in a path-sensitive manner, by maintaining at each point a distinct abstract fact (i.e., element of $L$) for distinct patterns of record types that could have been read so far. Typically,
to ``lift'' the analysis $U$ to a path-sensitive domain, a finite set of predicates of $P$ would be required~\cite{fischer2005joining}. The path-sensitive analysis domain would then essentially be $P \rightarrow L$. For our example program, a set of six predicates, each one formed by conjuncting one of the three predicates ``\pgm{in-rec.typ = `HDR'}'', ``\pgm{in-rec.typ = `ITM'}'' or ``\pgm{in-rec.type = `TRL'}'', with one of the two predicates ``\pgm{same-flag = 'S'}'' or ``\pgm{same-flag = 'D'}'', would be natural candidates. However, coming up with this set of predicates manually would be tedious, because it requires detailed knowledge of the state variables in the program and their usage. Automated predicate refinement~\cite{fischer2005joining} might be able to generate these predicates, but is a complex iterative process, and might potentially generate many additional predicates, which would increase the running time of the analysis.

\begin{figure}
\begin{minipage}{3.9in}
  \begin{tabular}{p{2.7in}p{1.2in}}
\begin{minipage}{2.7in}
\includegraphics[width=2.7in]{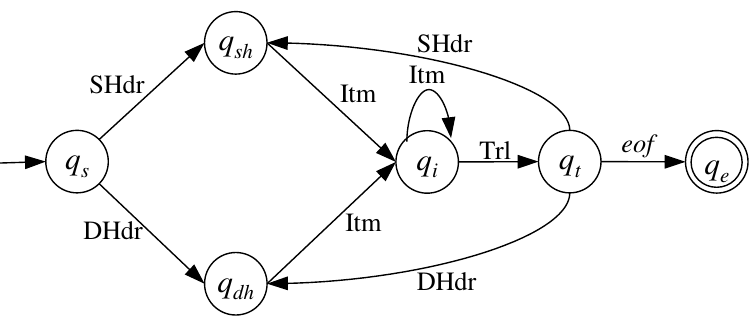}
\end{minipage}
&
\begin{minipage}{1.4in}
\begin{scriptsize}
\begin{tabular}{|c|l|}
\hline
Record  & \multicolumn{1}{c|}{Constraint}\\
 Type   & \\
\hline
SHdr       & \texttt{typ} = `{HDR}'$\wedge$
             \texttt{src}  = {`SAME'}\\
\hline
DHdr       & \texttt{typ} = `{HDR}'$\wedge$
            \texttt{src}  = `{DIFF}' \\
\hline
Itm        & \texttt{typ} = `{ITM}' \\
\hline
Trl        & \texttt{typ} = `{TRL}'\\
\hline
\end{tabular}
\end{scriptsize}

\end{minipage}
\\
\multicolumn{1}{c}{(a)} & \multicolumn{1}{c}{(b)}
\\
\end{tabular} 

\end{minipage}
 \caption{(a)  Well-formed input automaton. (b) Input record types. }
\label{fig:RunningOther}
\end{figure}

File-format specifications, which are usually readily available because they are organization-wide or even industry-wide standards, have been used by previous programming languages researchers in the context of tasks such as parser and validator generation~\cite{fisherPads2011}, and
white-box testing~\cite{godefroid2008grammar}. Our key insight is that if a file-format specification can be represented as a finite-state \emph{input automaton}, whose transitions are labeled with record types, then the set of states $Q$ of this automaton (which we call \emph{file states}) can be directly used to lift the analysis $U$, by using the domain $Q \rightarrow L$. The intuition is that if an abstract fact $l \in L$ is mapped to a file state $q \in Q$ at a program point, then $l$ over-approximates all possible concrete states that can arise at that point during runs that consume a sequence of records such that the concatenation of the types of these records is ``accepted'' by the file state $q$ of the automaton.

Figure~\ref{fig:RunningOther}(a) shows the well-formed input automaton for the file-format used in our running example, with Figure~\ref{fig:RunningOther}(b) showing the associated input record type descriptions (as \emph{dependent} types~\cite{XiDependent}). The sample input file in Figure~\ref{fig:RunningPgm}(b) is a well-formed file as per this automaton. This is because the sequence that consists of the types of the records in this file, namely `SHdr Itm Itm Itm Trl DHdr Itm Itm Trl', 
is accepted by the automaton.

Intuitively, statements other than \pgm{READ} statements do not affect the file-state that a program is ``in'' during execution. Therefore, the ``lifted'' transfer functions for these are straightforward, and use the underlying the transfer functions from the $L$ analysis. The transfer function for \pgm{READ} plays the key role of enforcing the ordering among record types in well-formed files. For instance, consider the file-state $q_\mathit{sh}$ in Figure~\ref{fig:RunningOther}(a), which represents the situation wherein a `\pgm{SAME}'-type header record has just been read. Therefore, in the output of the \pgm{READ} transfer function, $q_\mathit{sh}$ is mapped to the \emph{join} of the abstract facts that the predecessors file-states of $q_\mathit{sh}$, namely, $q_s$ and $q_t$ were mapped to in the input to the transfer function.

\paragraph{Applications.} 
In addition to our basic approach above, we propose two applications of it that address two natural problems in the analysis of file-processing programs, that to our knowledge have not been explored previously in the literature. The first application is a sound approach to check if a program potentially ``over accepts'' ill-formed files, or ``under accepts'' well-formed files. The second is a sound technique to specialize a program wrt a given specialization criterion that represents a restriction of the full file-format, and that is itself represented as an input automaton. 

\paragraph{Program File State Graph (PFSG).}
We propose a novel program
representation, the PFSG, which is a graph derived from both the
control-flow graph (CFG) of the program and the given \emph{input automaton}
for the program.  The PFSG is basically an ``exploded'' version of the CFG of the original program; the control-flow paths in the PFSG are a subset of the control-flow paths in the CFG, with certain paths that are infeasible under the given input automaton being omitted. Being itself a CFG, any existing static analysis can be applied on the PFSG without any modifications, with the benefit that the infeasible paths end up being ignored by the analysis.

We describe how to modify our basic approach to emit a PFSG, and also discuss formally how the results from any analysis differ when performed on the PFSG when compared to being performed on the original CFG.

\paragraph{Implementation and empirical results.}
We have implemented our approach, and applied it on several realistic as well as real Cobol batch programs.  Our approach found  file-format related conformance issues in certain real programs, and was also able to verify the absence of such errors in other programs. In the program specialization context, we observed that our approach was surprisingly precise in being able to identify statements and conditionals that need not be retained in the specialized program. We found that our analysis, when used to identify references to possibly uninitialized variables and reaching definitions gave improved precision over the standard analysis in many cases.\\[1em]

The rest of this paper is structured as follows. In Section~\ref{sec:FileFormat} we introduce key assumptions and definitions. In Section~\ref{sec:FileStateAnalysis} we present our approach, as well as the two applications mentioned above. Section~\ref{sec:pfsg} introduces the PFSG.  Section~\ref{sec:Implementation} presents our implementation and result. Section~\ref{sec:Related} discusses related work, while Section~\ref{sec:concl-future-work} concludes the paper.

\section{Assumptions and definitions}
\label{sec:FileFormat} 

\paragraph{Definitions (Records, Record Types, and Files)} A
\emph{record} is a contiguous sequence of bytes in a file.  A \emph{field}
is a labeled non-empty sub-string of a record. Any record has zero or more
fields (if it has zero fields then the record is taken to be a leaf-level
record). 

A \emph{record type} $R_i$
is intuitively a specification of the length of a record, the names of its
fields and their lengths, and a
\emph{constraint} on the contents of the record.
For example, consider the record types shown in
Figure~\ref{fig:RunningOther}(b).  Each row shows the name of a record
type, and then the associated constraint.

We say that a record $r$ \emph{is of type} $R_i$ iff $r$ satisfies the length
as well as value constraints of type $R_i$. For instance, the first record
in the file in Figure~\ref{fig:RunningPgm}(b) is of type SHdr (see
Figure~\ref{fig:RunningOther}(b)).  Note that in general a record $r$ could
be of multiple types.

\paragraph{Definitions (Files and read operations)}
A file is a sequence of records, of possibly different lengths.  Successive
records in a file are assumed to be demarcated explicitly, either by
inter-record markers or by other meta-data that captures the length of each
record. At run time there is a \emph{file pointer} associated with each
open file;
a \pgm{READ} statement, upon execution, retrieves the record pointed to by
this pointer, copies
it into the file buffer in the program associated with this file, and
advances the file pointer.

\paragraph{Definition (Input automaton)}

An \emph{input automaton} $S$ is a tuple $(Q, \Sigma, \Delta, q_s, Q_e)$,
where $Q$ is a finite set of states, which we refer to as file states,
$\Sigma$ = $\mathcal{T} \cup \{\eof\}$, where $\mathcal{T}$ is a set of
record types, $\Delta$ is a set of transitions between the file states,
with each transition labeled with an element of $\Sigma$, $q_s$ is the
designated \emph{start} state of $S$, and $Q_e$ is the (non-empty) set of
designated \emph{final} states of $S$.  A transition is labeled with $\eof$
iff the transition is to a final state. There are no outgoing transitions
from final states.

Note that an input automaton may be non-deterministic, in two different
senses.  Multiple transitions
out of a file state could have the same label. Also, it is possible for a
record $r$ to be of two distinct types $t_1$ and $t_2$ and for these two
types to be the labels of two outgoing transitions from a file state.

Let $q$ be any non-final state of $Q$.  We define $L_T(q)$ -- the
\emph{type language} of $q$ -- as the set of sequences of types (i.e.,
elements of $\mathcal{T}^*$) that take the automaton from its start
state to $q$. For a final state $q_e$, $L_T(q_e)$ is defined to be the
union of type languages of the states from which there are transitions to
$q_e$.

We define $L_R(q)$ -- the \emph{record language} of any file-state $q$ --
as follows: $L_R(q)$ consists of sequences of records $R$ such that there
exists a sequence of types $T$ in $L_T(q)$ such that (a) the sequences $R$
and $T$ are of equal length, and (b) for each $1 \leq j \leq |R|$, record
$R[j]$ is of type $T[j]$.  Recall that a file is nothing but a sequence of
records. We say that a file $f$ \emph{conforms} to an input automaton $S$, or
that $S$ \emph{accepts} $f$, if
$f$ is in $L_R(q_e)$ for some final state $q_e$ of $S$.

Let $R$ be some sequence of records (possibly empty). Say $R$ is in
$L_R(q)$ for some file state $q$ of an input automaton.  If there exists an
execution trace $t$ of the given program $P$ that starts at the program's
entry, consumes the records in $R$ via the \pgm{READ} statements that it
passes through, and reaches a program point $p$, then we say that (a) trace
$t$ is \emph{due to} the prefix $R$, and (b) trace $t$ reaches point $p$
while being \emph{in} file-state $q$ of the input automaton. We define
$\ns{t}$ as the sequence of nodes of the control-flow graph (CFG) of given
program that are visited by the trace $t$; the sequence always begins with
the entry node of the CFG and contains at least two nodes (the trace ends
at the point \emph{before} the last node in the sequence).

A well-formed input automaton (which we often abbreviate to ``well-formed
automaton'') is an input automaton that accepts all files that are expected
to be given as input to a program. A ``specialization'' automaton is an
input automaton that accepts a subset of files as a well-formed automaton,
while a ``full'' automaton is an input automaton that accepts \emph{every
  possible} file.

If a program accesses multiple sequential input files this situation could
be handled using two alternative approaches: (1) By concurrently using
multiple input automatons in the analysis, one per input file, or (2) By
modeling one of the input files as the primary input file (with an
associated automaton) and by modeling reads from the remaining files as
always returning an undefined record. We adopt the latter of these
approaches in our experimental evaluation.

\section{Our approach}
\label{sec:FileStateAnalysis}

In this section we describe our primary contribution, which is a generic approach for ``lifting'' a given abstract interpretation wrt a file-format specification. We then discuss its soundness and precision. Following this
we present the details of the two applications of our generic approach that were mentioned in Section~\ref{ssec:intro:approach}.
Finally, we present an extension to our approach, which enables the  specification of data integrity constraints on the contents of input files in relation to the contents of persistent tables.

\subsection{Abstract interpretation lifted using input automatons}
\label{ssec:approach:abstr-interpr-lift}
The inputs to our approach are a program $P$, an input automaton $S$ = $(Q, \Sigma, \Delta, q_s, Q_e)$, and an arbitrary ``underlying'' abstract interpretation $U$ $\equiv$ $((L, \sqsubseteq_L), F_L)$, where $L$ is a join semi-lattice and $F_L$ is a set of transfer functions with signature $L \rightarrow L$ associated with statements and conditionals. Our objective, as described in the introduction, is to use the provided input automaton to compute a least fix-point solution considering only paths in the program that are potentially feasible wrt the given input automaton.

The lattice that we use in our lifted analysis is $D$ $\equiv$ $Q \rightarrow L$. The partial ordering for this lattice is a ``point wise'' ordering based on the underlying lattice $L$:
$$ d_1 \sqsubseteq_D d_2 \ =_\mathit{def} \ \forall q \in Q. \, d_1(q) \sqsubseteq_L d_2(q) $$

The initial value that we supply at the entry of the program is $(q_s, i_L)$, where $i_L \in L$ is an input to our approach, and is the initial value to be used in the context of the underlying analysis. 

We now discuss our transfer functions on the lattice $D$. We consider the following three categories of CFG nodes: Statements \emph{other} than \pgm{READ} statements, conditionals, and \pgm{READ} statements. Let $n$ be any node that is neither a \pgm{READ} statement nor a conditional. Let $f^n_L:L \rightarrow L$ $\ \in\ $ $F_L$ be the ``underlying'' transfer function
for node $n$. Since the file state that any trace is \emph{in} at the point before node $n$ cannot change after the trace executes node $n$,  our transfer function for node $n$ is:
$$f^n_D(d\in D) \ = \ \lambda q \in Q.\, f^n_L(d(q))$$

Let $c$ be a conditional node, with a {\tru} successor and a {\fls}
successor. Let $f^c_{t,L} \in F_L$ and $f^c_{f,L} \in F_L$ be the underlying {\tru}-branch and
{\fls}-branch transfer functions of $c$. Since a conditional node
cannot modify the file-state that a trace is in, either, our transfer function for
conditionals is:
$$f^c_{b,D}(d\in D) = \lambda q \in Q.\, f^c_{b,L}(d(q))$$

where `$b$' stands for $t$ or $f$.

\begin{figure}
\centering
\input{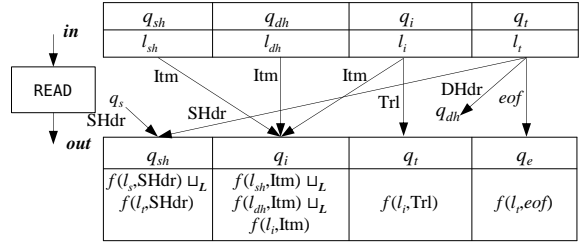}
\caption{Illustration of 
  transfer function for \pgm{READ} statements.}
\label{fig:xfer-funcs}
\end{figure}

Finally, we consider the case where a node $r$ is a \pgm{READ} node.
This is
the most interesting case, because executing a \pgm{READ} statement can change the file state that a trace is in.
Firstly, a note on  terminology: a dataflow value $l \in L$ is said to \emph{represent} a concrete state $s$ if $s$ is an element of the \emph{concretization}~\cite{Cousot77} of $l$ (which is written as $\gamma(l)$). Secondly, we make the following assumption on the underlying transfer function $f^r_L \in F_L$ for \pgm{READ} statements:
Rather than simply have the signature $L \rightarrow L$, the function $f^r_L$ ought to have the signature $(L \times \Sigma) \rightarrow L$.  If $t$ is some record type (i.e., element of $\mathcal{T}$), then, intuitively, $f^r_L(l_1,t)$ should return a dataflow fact $l_2$
that represents the set of concrete states that can result after the execution of the \pgm{READ}, assuming:

\begin{itemize}
\item the concrete state just before the execution of the \pgm{READ} is some state that is represented by $l_1$, and
\item the \pgm{READ} statement retrieves a record of type $t$ from the input file and places it in the input buffer.
\end{itemize}

Correspondingly, $f^r_L(l_1, \eof)$ should return a dataflow fact $l_2$ that represents the set of concrete states that can result after the execution of the \pgm{READ}, assuming:
\begin{itemize}
\item the concrete state just before the execution of the \pgm{READ} is some state that is represented by $l_1$, and
\item the input buffer in the program gets populated with an undefined value, and
\item the statement within the `\pgm{AT END}' clause, if any, executes after the read operation.
\end{itemize}

As an illustration, say the underlying analysis $U$ is the CP (Constant Propagation) analysis.  $f^r_L(l_1, t)$ would return a fact $l_2$ that is obtained by performing the following transformations on $l_1$: (1) remove all existing CP facts associated with the input buffer, and (2) obtain suitable new CP facts for the input buffer using the \emph{constraints} associated with the type $t$. On the other hand, $f^r_L(l_1, \eof)$ would perform only Step~(1) above.
\hfill$\Box$

We are now ready to present our transfer $f^r_D$ for a read node $r$. The transfer function is:
$$f^r_D(d) \ = \ \lambda q_j \in Q.\, \bigsqcup_{(q_i
  \rightarrow q_j) \in \Delta} \{ f^r_L(d(q_i), \mathit{label}(S,q_i,q_j))\}$$

where \emph{label}($S,q_i,q_j$) returns the label (which is a type, or $\eof$) of the transition $q_i \rightarrow q_j$ in $S$. The intuition behind this transfer function is as follows. For any file state $q_j$, a trace can be in $q_j$ after executing the \pgm{READ} if the trace is in any one of the predecessor states of $q_j$ in the automaton just before executing the \pgm{READ}. Therefore, the fact (from lattice $L$) that is to be associated with $q_j$ at the point after the \pgm{READ} statement can be obtained as follows: (1) For each file state $q_i$ such that there is a transition $q_i \rightarrow q_j$ labeled $s$ in the automaton, transfer the fact $f^r_L(l_1,s)$, where $l_1 \in L$ is the fact that $q_i$ is mapped to at the point before the \pgm{READ} statement. (2) Take a join of all these transferred facts.

Figure~\ref{fig:xfer-funcs} sketches this transfer function schematically. Each edge from a column before the \pgm{READ} statement to a column after the \pgm{READ} denotes a ``transfer'' that happens due to Step~(1) above; the label on the edge denotes the label on the corresponding transition in the automaton. We have abbreviated each instance of $f^r_L$ in this figure as $f$.  We have also omitted some of the columns for compactness.

Our presentation above was limited to the intra-procedural setting. However,  our analysis can be extended to the inter-procedural setting using standard techniques, some details of which we discuss   in Section~\ref{sec:Implementation}. 

\subsubsection{Illustration.}

\begin{figure}[t]
  \centering
  \includegraphics{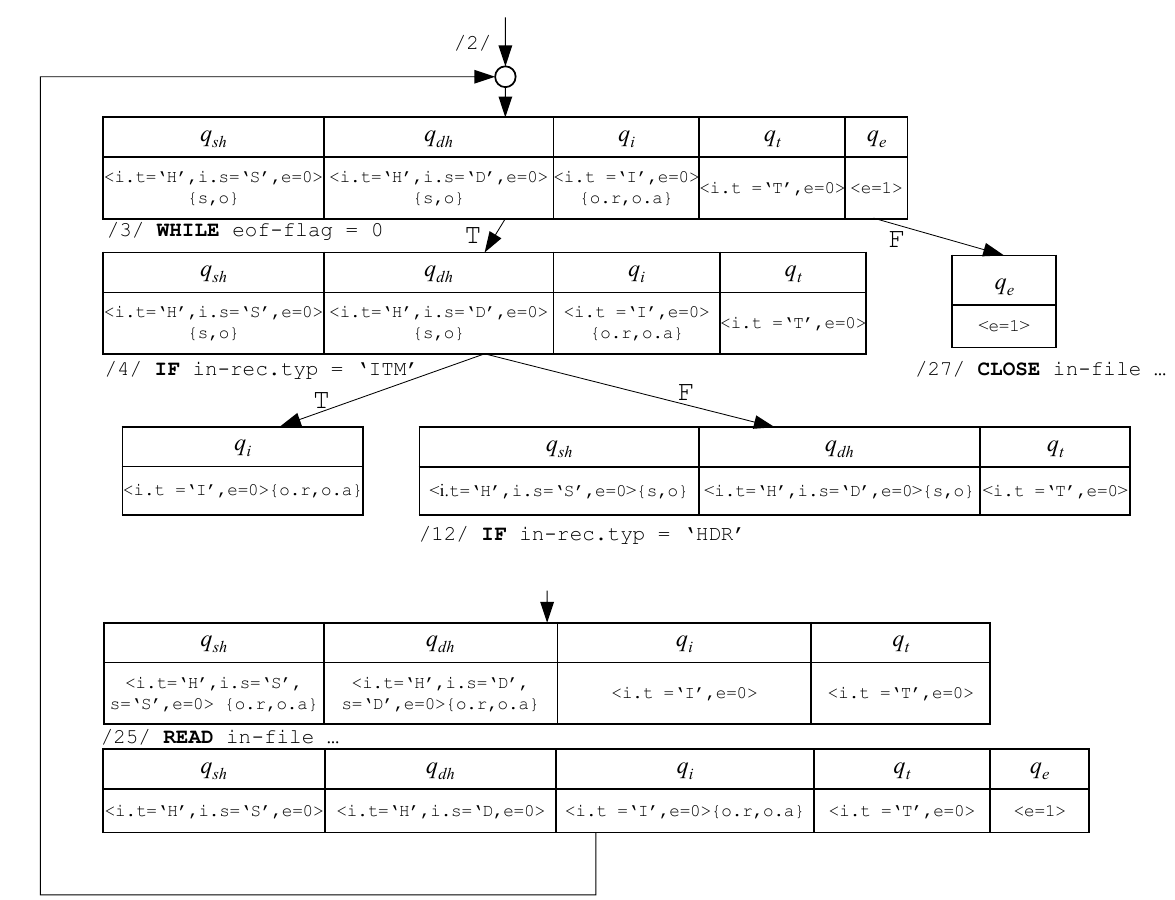}
     \caption{Fix point solution for program in Figure~\ref{fig:RunningPgm}(a),  using CP $\times$ Possibly-unitialized-variables analysis. Abbreviations used: \pgm{e}: \pgm{eof-flag}, \pgm{s}: \pgm{same-flag}, \pgm{i}: \pgm{in-rec}, \pgm{o}: \pgm{out-rec}, \pgm{i.t}: \pgm{in-rec.typ}, \pgm{i.s}: \pgm{in-rec.src}, \pgm{o.r}: \pgm{out-rec.rcv}, \pgm{o.a}: \pgm{out-rec.amt}, 'H': 'HDR', 'S': 'SAME', 'D': 'DIFF', 'I;: 'ITM', 'T': 'TRL'.}

  \label{fig:RunningSoln}
\end{figure}

Figure~\ref{fig:RunningSoln} shows the fix-point solution at certain program points for the example program in Figure~\ref{fig:RunningPgm} as computed by our analysis, using the well-formed automaton shown in Figure~\ref{fig:RunningOther}. We assume an underlying lattice $L$ that is a product of the constant-propagation (CP) and possibly uninitialized variables (Uninit) lattices.
Each table in the figure denotes the solution (i.e., a function from $Q$ to $L$) at the program point that precedes the statement that is indicated below the table. Each column of a table shows the underlying dataflow value associated with a  file state. Columns in which the underlying dataflow value is $\bot$ (which represents unreachability, basically) are omitted from the tables for brevity. The first component of each dataflow value -- within angle brackets -- indicates the constant values of variables, while the second component -- within curly braces --  indicates the set of variables that are possibly uninitialized. Empty sets are omitted from the figure for brevity. We abbreviate the variable names as well as constant values  for the sake of compactness, as described in the caption of the figure.

In the interest of space, we focus our attention on just one of the program points -- the point just before the \pgm{IF} condition in line~4. Any execution trace reaching this point can be \emph{in} any one of the following four file states: $\qsh, \qdh, q_i$, or $q_t$. These file states have  associated CP facts that indicate that \pgm{in-rec.typ} has value  `\pgm{HDR}', `\pgm{HDR}', `\pgm{ITM}',  and `\pgm{TRL}', respectively. Additionally, the state variable \pgm{same-flag} is possibly uninitialized under columns $\qsh$ and $\qdh$, because lines~14-17 (which initialize this variable) may not have been visited yet, whereas is initialized under the other two columns. Now, only the fact associated with $q_i$ flows down the \emph{true} branch of the conditional in line~4. This is because this conditional tests that \pgm{in-rec.typ} contains `\pgm{ITM}'. Therefore, \pgm{same-flag} is inferred to be definitely initialized by the time it is referenced in line~6, which is the desired precise result.

\subsection{Soundness, precision, and complexity of our approach}
\label{ssec:approach:soundn-prec-compl}

Our soundness result, intuitively, is that if the underlying analysis $U$ is sound, then so is our lifted analysis, modulo the assumption that any input file given to the program $P$ conforms to the given input automaton $S$. $U$ itself is said to be sound if at any program point $p$ of any program $P$ the fix-point solution $l$ computed by $U$ represents all concrete states that can result at $p$ due to \emph{all} possible execution traces of $P$ that begin in any concrete state that is represented by the given initial value $i_L$. 
The following theorem states the soundness characterization of our analysis more formally.

\begin{theorem}\label{corr:soundness}
Assuming $U$ is a sound abstract interpretation, if $d$ is the fix-point solution produced by our approach at a program point $p$ of the given program $P$ starting with initial value $(q_s, i_L)$, then $l$ $\equiv$ $\sqcup_{q \in Q} \{d(q)\}$ represents all concrete states that can result at point $p$ due to all possible executions that begin in any concrete state that is represented by $i_L$ and that are on an input file that conforms to the given automaton $S$. 
\end{theorem}

The proof of the theorem above is   straightforward. 

The following observations follow from the theorem above. (1) If the given input automaton is a well-formed automaton, then the fix-point solution at a program point $p$ represents all concrete states that can result at point $p$ during executions on well-formed files. (2) If the given input automaton is a ``full'' automaton, then the fix-point solution at a program point $p$ represents all concrete states that can result at point $p$ during all possible executions (including on ill-formed files). 

Given \emph{any} input automaton $S$, our approach will produce a fix-point solution that is at least as precise as the one that would be produced directly by the underlying analysis $U$. However, the choice of the input automaton does impact the precision of the our approach. Intuitively, if input automaton $S_1$  is a sub-automaton of input automaton $S_2$ (i.e., is obtained by deleting certain states and transitions, or by constraining further some of the types that label some of the transitions), then $S_1$ will result in a more precise solution than $S_2$. Also, if $S_1$ and $S_2$ accept the same language, but $S_1$ structurally \emph{refines} $S_2$, then $S_1$ will result in a more precise solution. We formalize these notions in the appendix.

The time complexity of our analysis when used with an automaton $S$ that has a set of states $Q$ is, in the worst case, $|Q|$ times the
worst-case time complexity of the underlying analysis $U$.

\subsection{Two applications of our analysis}
\label{sec:Applications}

In this section we describe how we use our analysis described in Section~\ref{ssec:approach:abstr-interpr-lift} to address the two new problems that we mentioned in the introduction -- file format conformance checking, and  program specialization.

\subsubsection{File format conformance checking.}
\label{ssec:appl:conformance}

As mentioned in the introduction, a verification question that developers would like an answer to is whether a program can silently ``accept'' an ill-formed input file and possibly write out  a corrupted output file (``over acceptance''). Or, conversely, could the program ``reject''  a well-formed file via an abort or a warning message (``under acceptance'')?

Different programs use different kinds of idioms to ``reject'' an input file; e.g., generating a warning message (and then continuing processing as usual), ignoring an erroneous part of the input file and processing the remaining records, and aborting the program via an exception. In order to target all these modes in a generic manner, our approach relies on the developer to identify file-format related \emph{rejection points} in a program. These are the statements in a program where format violations are flagged, using warnings, aborts, etc.

\paragraph{Detecting under-acceptance.}

We detect under-acceptance warnings by (1) Applying our analysis using a well-formed automaton and using any given program-state abstraction domain (e.g., CP, or \emph{interval analysis}) as $U$.  (2) Issuing an under-acceptance warning if the fact  associated with any file-state is non-$\bot$ at any rejection point. The intuition is simply that rejection points should be unreachable when the program is run on well-formed files. Since our analysis is conservative, in that it never produces under-approximated dataflow facts, this approach will not miss any under-acceptance issues as long as the developer does not fail to mark any actual rejection point as a rejection point. 

As an illustration, say line~23 in the program in Figure~\ref{fig:RunningPgm} is marked as a rejection point. Using the well-formed automaton in Figure~\ref{fig:RunningOther}(a) and using CP as the underlying analysis $U$, our analysis will find this line to be unreachable. Therefore, no under-acceptance warnings will be issued.

\paragraph{Detecting over-acceptance.}

Intuitively, a program has \emph{over acceptance} errors with respect to a given well-formed file format if the program can reach the end of the main procedure without going through any rejection point when run on an input file that \emph{does not} conform to the well-formed automaton. We check this property as follows: We first \emph{extend} the given well-formed automaton $S$ to a \emph{full} automaton (which accepts \emph{all} input files) systematically by adding a new final state $q_x$, a few other new non-final states,  and new transitions that lead to these new states from the original states. The intent is for these new states to accept record sequences that are not accepted by any file state in the original automaton $S$.   We provide the full details of this construction in the appendix. Secondly, we modify the transfer functions of our lifted analysis $D$ at rejection points such that they map \emph{all} file states to  $\bot$ in their output. Intuitively, the idea behind this is to ``block'' paths that go through rejection points.

We then apply our analysis using this full automaton and using any program-state abstraction domain as $U$, and flag an over-acceptance warning if the dataflow
value associated with any file state that is not a final state in the original well-formed automaton is
non-$\bot$ at the final point of the ``main'' procedure. Clearly, since our analysis over-approximates dataflow facts at all program points, we will not miss any over-acceptance scenarios as long as the developer does not wrongly mark a non-rejection point as a rejection point.

\subsubsection{Program specialization based on file formats.}
\label{ssec:appl:spec}
As mentioned in the introduction, it would be natural for developers to want to specify specialization criteria for file-processing programs as patterns on sequences of record types in an input file. We propose the use of input automatons for this purpose. For example, if the well-formed automaton in Figure~\ref{fig:RunningOther}(a) were to be modified by removing the file state $\qdh$, as well as all transitions incident on it, what would be obtained would be a \emph{specialization automaton} that accepts files in which all batches begin with ``same'' headers only. 

Our approach to program specialization using a specialization automaton is as follows. (1) We apply our analysis of Section~\ref{ssec:approach:abstr-interpr-lift} using the given specialization automaton as the input automaton, and using
any program-state abstraction domain as $U$.
(2) We identify program points $p$ at which every file state is mapped to $\bot$ as per the fix-point solution computed in Step 1. Basically, these program points are unreachable during executions on input files that conform to the specialization criterion. The statements/conditionals that immediately follow these points can be projected out of the program to yield the specialized program (the details of this projection operation are not a focus of this paper).

It is easy to see that our approach is sound, in that it marks a point as unreachable only if it is definitely unreachable during all runs on input files that adhere to the given criterion.

\paragraph{Illustration.} Using the specialization automaton mentioned above, and using CP as the underlying analysis, lines~9 and~17 in the code in Figure~\ref{fig:RunningPgm}(a) are marked as unreachable. It is worthwhile noting that if one did not use the specialization automaton as criterion, and instead simply specified that all header records have value `\pgm{SAME}' in their `\pgm{src}' field, then line~9 would \emph{not} be identified as unreachable. Intuuitively, this is because the path consisting of lines~1-6, along which \pgm{same-flag} is uninitialized, would not be found infeasible,  as discussed in Section~\ref{ssec:analysis-issues}. 

Subsequently, as a post-processing step (which is not a part of our core approach), the following further simplifications could be done to the program: (i) Make lines~7 and~15  unconditional, and remove the respective controlling ``if'' conditions. This would be safe because the ``else'' branches of these two ``if'' conditions have become empty. (ii) Remove line~15 entirely. This would be safe because after the conditional in line~6 is removed the variable \pgm{same-flag} is not used anywhere in the program.

\subsection{Imposing data integrity constraints on input files}
\label{ssec:approach:persist}

The core of our approach, which was discussed in Section~\ref{ssec:approach:abstr-interpr-lift}, used input automatons that constrain the sequences of types of records that can appear in an input file. However, in many situations, a well-formed file also needs to satisfy certain data integrity constraints wrt the contents of certain persistent tables. If these constraints can also be specified in conjunction with the input automaton, then certain paths in the program that execute only upon the violation of these constraints can be identified and pruned out during analysis time. This has the potential to further improve the precision and usefulness of our approach.

In our running example,  say there is a requirement that the ``receiver'' of any payment  in the input file  (represented by the \pgm{in-rec.rcv} field in the item record) necessarily be an account holder in the bank. Such a requirement could be enforced in the code in Figure~\ref{fig:RunningPgm}  by adding logic right after line~4 to check if the value in \pgm{in-rec.rcv} appears as a primary key in the ``accounts'' database table of the bank, and to not execute lines~5-11 if the check fails. However, if a user of our approach wishes to assert that input files will never contain items that refer to invalid account numbers, then the logic mentioned above could be identified as redundant. To enable users to give such specifications  we allow predicates of the form \textit{isInTable}(\emph{Tab},\emph{field}) and \emph{isNotInTable}(\emph{Tab},\emph{field}) to be associated with record type definitions, where \emph{Tab} is the name of a persistent table, and \emph{field} is the name of a field in the record type. For example, the Itm type in Figure~\ref{fig:RunningOther}(b) could be augmented as ``\texttt{typ} = `\pgm{ITM}' $\wedge$ \emph{isInTable}(\pgm{accounts},\texttt{rcv})'', where \pgm{accounts} is the master accounts table. The semantics of this is that the value in the \pgm{rcv} field is guaranteed to be a primary key of some row in the table \pgm{accounts}. Similarly, \emph{isNotInTable}(\emph{Tab},\emph{field}) asserts that the value in \emph{field} is guaranteed \emph{not} to be a primary key of any row in \emph{Tab}.

We assume our programming language has the following construct for \emph{key-based lookup} into a table \emph{Tab}:

\texttt{READ} \emph{Tab} \texttt{INTO} \emph{buffer} \texttt{KEY} \emph{variable}, 
\texttt{INVALID KEY} \emph{statements-N}
\emph{statements-F}

\noindent The semantics of this statement is as follows:  If a table row with a key matching the value in \emph{variable} is found in the table \emph{Tab}, then it will be copied into \emph{buffer} and control is given to \emph{statements-F}. If no matching key is found, then the \emph{buffer} content is undefined and control is given to \emph{statements-N}.

With this enhancement of record type specifications, we extend our analysis framework as follows. The new lattice we use will be $D$ $\equiv$ $Q \rightarrow (S \times L)$ where $L$ is the given original underlying lattice, $S=2^{C}$ and $C$  is the set of all possible predicates of the two kinds mentioned above.

We now describe the changes required to the transfer functions.
The transfer function of (normal) \pgm{READ} statements that read from input files that we described in Section~\ref{ssec:approach:abstr-interpr-lift} is to be augmented, as follows. Whenever a record of a certain type $t \in \mathcal{T}$ is read in, any predicates in the incoming fact that refer to fields of the input buffer are removed, and the predicates associated with $t$ are included in the outgoing fact.

The transfer function of the statement ``\pgm{MOVE X TO Y}'' copies to the outgoing fact all predicates in the incoming fact that refer to variables other than \pgm{Y}. Further, for each predicate in the incoming fact that refers to \pgm{X}, it creates a copy of this predicate, makes it refer to \pgm{Y} instead of \pgm{X}, and adds it to the outgoing fact.

Transfer functions of conditionals do not need any change.

Finally, we need to handle key-based lookups, which is the most interesting case. Consider once again the statement:

\texttt{READ} \emph{T} \texttt{INTO} \emph{buffer} \texttt{KEY} \emph{v}, 
\texttt{INVALID KEY} \emph{statements-N}
\emph{statements-F}

The transfer function first checks if a predicate of the form \textit{isInTable}(\emph{T},\emph{v}) is present in the incoming dataflow fact. If it does, it essentially treats \emph{statements-N} as unreachable. Else, if a predicate of the form \textit{isNotInTable}(\emph{T},\emph{v}) is present in the incoming fact, it essentially treats \emph{statements-F} as unreachable. Otherwise, it treats both \emph{statements-F} and \emph{statements-N} as  reachable.

A more formal presentation of these transfer functions is omitted from this paper in the interest of space.

\section{The Program File State Graph (PFSG)}
\label{sec:pfsg}

In this section we introduce our program representation for file-processing
programs, the Program File State Graph (PFSG).  We then formalize the
properties of the PFSG. Finally, we discuss how the PFSG serves as a basis
for performing other program analyses \emph{without any modifications},
while enabling them to ignore certain CFG paths that are infeasible as per
the given input automaton.

\subsection{Structure and construction of the PFSG}
\label{ssec:pfsg-constr}

The PFSG is a representation that is based on a CFG $G$ of a
file-processing program $P$ as well as on a given input automaton $S$ for
$P$.

The PFSG is basically an \emph{exploded} CFG. If the set of states in $S$
(i.e., \emph{file states}) is $q_s, q_1, q_2, \ldots, q_e$, then, for each
node $m$ in the CFG, we have nodes $(m,{q_s})$, $(m,{q_1})$, $(m,q_2),
\ldots, (m,{q_e})$ in the PFSG.  In other words, the PFSG has $N |Q|$
nodes, where $N$ is the number of nodes in $G$ and $Q$ is the set of
file-states of $S$.  A structural property of the PFSG is that an edge is
present between nodes $(m,{q_i})$ and $(n,{q_j})$ in the PFSG \emph{only
  if} there is an edge from $m$ to $n$ in the CFG. In other words, any path
$(m,{q_i}) \rightarrow (n,{q_j}) \rightarrow \ldots \rightarrow (r,{q_l})$
in the PFSG \emph{corresponds to} a path $m \rightarrow n \rightarrow
\ldots \rightarrow r$ in the CFG. Let $s_G$ be the entry node of the
CFG. The node $(s_G,q_s)$ is regarded as the entry node of the PFSG.

The PFSG can be constructed in a straightforward manner using our basic
approach that was described in
Section~\ref{ssec:approach:abstr-interpr-lift}. The precision of the PFSG
is linked to the precision of the underlying analysis $U$ that is
selected. For example, the standard CP (\emph{constant-propagation})
analysis could be used as $U$.  If more
precision is required a more powerful lattice, then, for instance, a
\emph{relational} domain (wherein each lattice value represents a
\emph{set} of possible valuations of variables), such as the Octagon
domain~\cite{octagon2006}, could be used. Once the fix-point solution is
obtained from the approach, edges are added to the PFSG as per the
following procedure.

\noindent For each edge $m \rightarrow n$ in the CFG:
\begin{description}
  \item{Rule 1,} applicable if $m$ is a ``read'' node: For each transition
    $q_i \rightarrow q_j$ in the automaton $S$, add an edge $(m,q_i)
    \rightarrow (n,q_j)$ in the PFSG.

  \item{Rule 2,} applicable if $m$ is not a ``read'' node: For each $q \in
    Q$ add an edge $(m,q) \rightarrow (n,q)$ in the PFSG.
\end{description}

In both the rules above we add an edge $(m,q_k) \rightarrow (n,q_l)$
\emph{only if} $d_m(q_k) \neq \bot_L$ and $d_n(q_l) \neq \bot_L$, where
$d_m$ and $d_n$ are the fix-point solutions at $m$ and $n$, respectively.
We follow this restriction because $d_m(q_k)$ (resp. $d_n(q_l)$) is $\bot$
only when there is no execution trace that can reach $m$ (resp. $n$) due to
a sequence of records that is in $L_R(q_k)$ (resp. $L_R(q_l)$).

The intuition behind the first rule above is that when a ``read'' statement
executes, it modifies the file-state that the program could ``be in'';
intuitively, this is a file-state that the input automaton could be in
were we to start simulating the automaton from $q_s$ when the program
starts executing, and transition to an appropriate target state upon the
execution of each ``read'' statement based on the type of the record read.
When executing a ``read'' statement a program could transition from a
file-state $q_i$ to  file-state $q_j$ only if such a transition is present
in the input automaton. 

The second rule above does not ``switch'' the file-state of the program,
because statements other than ``read'' statements affect the program's
internal state (i.e, valuation of variables) but not the file-state that
the program is in.

\subsection{Illustration of PFSG}
\label{ssec:pfsg-illus}

\begin{figure}
\centering
\includegraphics[width=3.5in]{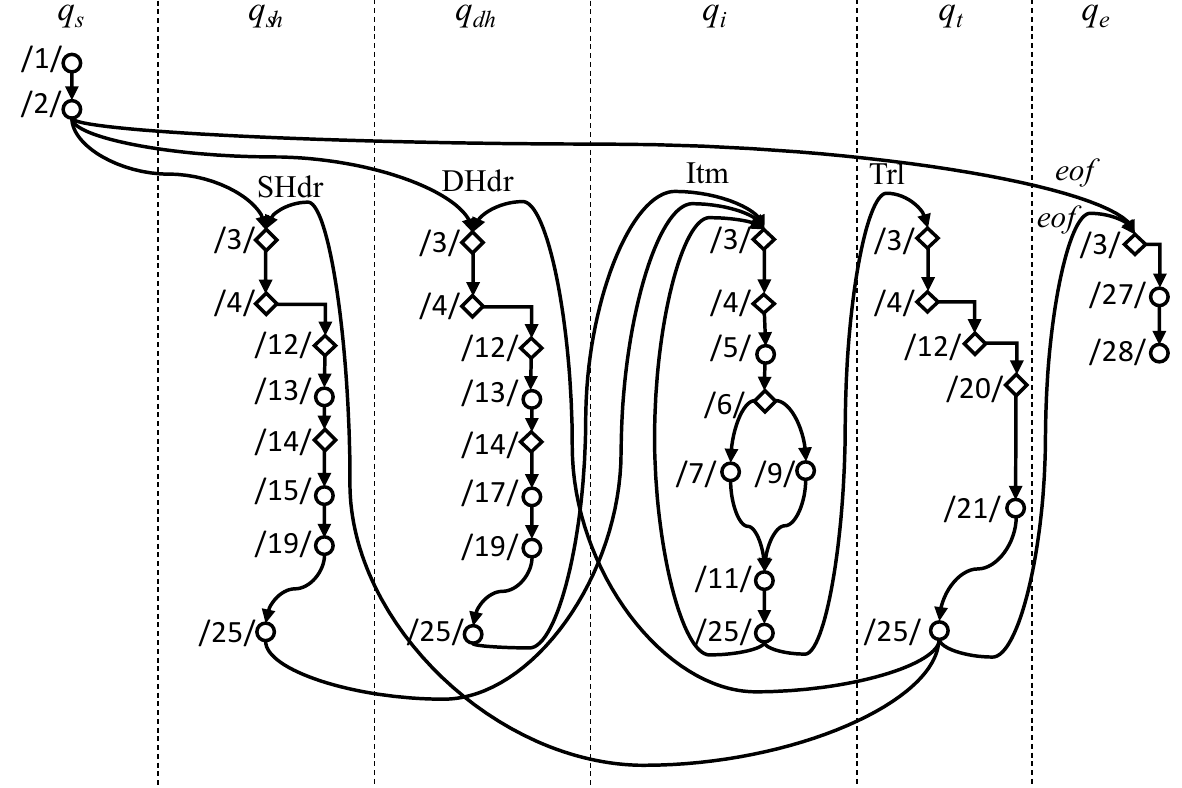}
  \caption{PFSG for program in Fig.~\ref{fig:RunningPgm}(a) and input
    automaton in Fig.~\ref{fig:RunningOther}}
  \label{fig:FullPFSG}
\end{figure}

For illustration, consider the PFSG shown in Figure~\ref{fig:FullPFSG},
corresponding to the program in Figure~\ref{fig:RunningPgm}(a) and input
automaton in Figure~\ref{fig:RunningOther}. Recall that this automaton
describes all well-formed files where headers, item records, and trailers
appear in their correct positions.  Visually, the figure is laid out in six
columns, corresponding to the six file states in the input automaton.  The
nodes in the PFSG are labeled with the corresponding line numbers from the
program. Therefore, e.g., the node labeled /1/ in the $q_s$ column is
actually node $(/1/,{q_s})$, where /1/ represents the \pgm{OPEN} statement
in line~1 of the program. On a related note, $q_s$ being the start state of
the input automaton, and line~1 being the entry node of the program, the
node mentioned above is in fact the entry node of the PFSG. Certain parts
of the PFSG are elided for brevity, and are represented using the cloud
patterns. This PFSG was generated using a fix-point solution from our
approach of Section~\ref{ssec:approach:abstr-interpr-lift}, using CP
(constant propagation) as the underlying analysis $U$. A fragment of this
fix-point solution was shown in Figure~\ref{fig:RunningSoln} (the sets of
possibly uninitialized variables in that figure can be ignored in the
current context.)

We now discuss in more detail a portion of the PFSG in
Figure~\ref{fig:FullPFSG}, with emphasis on how it elides certain
infeasible paths that are present in the original CFG.  Line~2 in the
program is a \pgm{READ} statement. As per the given input automaton the
outgoing transitions from $q_s$ go to $\qsh$ and to $\qdh$. Therefore, as
per Rule~1 of our PFSG edge-addition approach (see
Section~\ref{ssec:pfsg-constr} above), there are outgoing edges from
$(/2/,q_s)$ to copies of node~3 in the $\qsh$ and $\qdh$ columns (for
clarity we have labeled these edges with the types on the corresponding
input-automaton transitions). The $\qsh$ column (i.e., the second column)
essentially consists of a copy of the loop body, \emph{specialized} to the
situation wherein the last record read was of type SHdr (SHdr being the
type on all transitions coming into $\qsh$). In particular, note that the
{\tru} edge out of /4/ to /5/ in the $\qsh$ column is elided. This is
because in the fix-point solution (see Figure~\ref{fig:RunningSoln}) the CP
(constant propagation) fact associated with the $\qsh$ file state at the
point before /4/ indicates that \pgm{in-rec.typ} has value `\pgm{HDR}'
(this fact is abbreviated as ``\pgm{i.t = `H'}'' in the figure). Therefore,
in the fix-point solution, the underlying fact associated with the $\qsh$
file state out of this edge ends up being $\bot$, which results in Rule~2
adding only the {\fls} edge from /4/ to /12/. 

Line~25 in the program being a \pgm{READ} statement, there is an edge from
node $(/25/,\qsh)$ at the bottom of the $\qsh$ column to the entry of
column $q_i$ ($q_i$ being the sole successor of $\qsh$ in the input
automaton). The $q_i$ column consists of a copy of the loop body,
specialized to the situation wherein the previous record read is of type
Itm (this being the type on transitions coming into $q_i$).  From the end
of the $q_i$ column control goes to the $q_t$ column, and from the end of
that column back to the beginning of the $\qsh$ and $\qdh$ columns.

It is notable that the structure of the PFSG is inherited both from the CFG
and from the input automaton. As was mentioned in the discussion above,
control transfers from one column to another mirror the transitions in the
input automaton, while paths within a column are inherited from the CFG,
but specialized wrt the type of record that was last read.

\subsection{Program analysis using PFSG}

Any program analysis that can be performed using a CFG can naturally be
performed \emph{unmodified} using a PFSG, by simply letting the
analysis treat each node $(n,q_i)$ as being the same statement/conditional
as the underlying node $n$.

Such an analysis will be no less precise than with the original CFG of the
program. This is because, by construction, every path in the PFSG
corresponds to a path in the original CFG; in other words, there are no
``extra'' paths in the PFSG. To the contrary, certain CFG paths that are
infeasible as per the given input automaton could be omitted from the
PFSG. In other words, precision of the analysis is improved by ignoring
executions due to certain infeasible inputs.  For instance, in the example
that was discussed above, due to the omitted edge from /4/ to /5/ in the
$\qsh$ column, there is no path in the PFSG that visits copies of the nodes
/1/, /2/, /3/, /4/, /5/, and /6/, in that order, even though such a path
exists in the original CFG. In other words, the PFSG encodes the fact that
under the given input file format an ``item'' record cannot occur as the
first record in an input file. (However, in general, due to possible
imprecision in the given underlying analysis $U$, not all paths that are
infeasible as per the given input automaton would necessarily be excluded
from the PFSG.)

To illustrate the benefits of program analysis using the PFSG, we discuss
two example analyses below:

\begin{itemize}
\item Say we wish to perform possibly uninitialized variables analysis. Due
  to the path /1/-/2/-/3/-/4/-/5/-/6/ in the original CFG, the use of
  \pgm{same-flag} in line~/6/ would be declared as possibly
  uninitialized. However, under the given input automaton, since every path
  that reaches line~/6/ in the PFSG reaches it via lines~/15/ or~/17/
  (which both defined \pgm{same-flag}), the use mentioned above would be
  declared as definitely initialized when the possibly-initialized analysis
  is performed on the PFSG.

\item A CP (constant-propagation) analysis, when done on the PFSG in
  Figure~\ref{fig:FullPFSG}, would indicate that at the point before
  line~/25/ \pgm{same-flag} would not have a constant value. However, when
  the same analysis is done on the PFSG, the same analysis would indicate
  that if (a) if \pgm{in-rec.type} is `\pgm{HDR}' and \pgm{in-rec.src} is
  `\pgm{SAME}' then \pgm{same-flag} has value `\pgm{S}', (b) if
  \pgm{in-rec.type} is `\pgm{HDR}' and \pgm{in-rec.src} is `\pgm{DIFF}'
  then \pgm{same-flag} has value `\pgm{D}', and (c) \pgm{same-flag} is not
  a constant otherwise. These correlations are identified because the PFSG
  is ``exploded'', hence segregates CFG paths that end at the same program
  point but are due to record sequences that are accepted by different
  file-states of the input automaton. Correlations such as the one
  mentioned above cannot be identified, in general, using the CFG unless
  very expensive domains (such as relational domains) are used.
\end{itemize}

The two instances of precision improvement mentioned above can also be
obtained using our approach of
Section~\ref{ssec:approach:abstr-interpr-lift}, if we use \mbox{CP $\times$
  Uninit} as the underlying domain $U$ (where Uninit is the
possibly-uninitialized analysis) for the first instance, and if we simply
use CP as the underlying domain $U$ for the second instance. However, in
general, there are several scenarios where the PFSG serves better as a
foundation for performing program analysis than the approach of
Section~\ref{ssec:approach:abstr-interpr-lift}:

\begin{itemize}
\item The approach of Section~\ref{ssec:approach:abstr-interpr-lift}
  applies only to \emph{forward} dataflow analysis. Whereas, the PFSG can
  be used for forward as well as backward dataflow analysis problems.

\item The PFSG as a basis for applying static analysis techniques other
  than dataflow analysis, such as symbolic execution, model-checking,
  assertional reasoning, etc. Implementations of these techniques that are
  designed for CFGs can be applied \emph{unmodified} on the PFSG.
  All these analyses are
  likely to benefit from the pruning of paths from the PFSG that are
  infeasible as per the given input automaton.
\end{itemize}

\subsection{Formal properties of the PFSG}

\subsubsection{Soundness.}
We now characterize the paths in the original CFG that are necessarily
present in the PFSG. This result forms the basis for the soundness of any
static analysis that is applied on the PFSG.\\
 
\emph{Theorem:} Let $U$ $\equiv$ $((L, \sqsubseteq_L), F_L)$ be a given
underlying \emph{sound}~\cite{Cousot77} abstract interpretation. Consider
any execution trace $t$ of the program $P$ that begins with a concrete
state that is \emph{represented} by the given initial dataflow fact $i_L
\in L$. Let $l$ be the sequence of records \emph{due to} which $t$
executes, and $T$ be the number of nodes in $\ns{t}$.

\noindent \textbf{If}

(a) $l$ is in $L_R(q)$ for some non-final file-state $q$ of $S$, and $t$ did
not encounter end-of-file upon a read, \emph{or}

(b) $l$ is in $L_R(q)$ for some final file-state $q$ of $S$, and the last
``read'' in $t$ encountered end-of-file

\noindent \textbf{Then} there is a path $t'$ in the PFSG such that

(a) the first node of $t'$ is $(s_G,q_s)$,

(b) for all $i \in [2,T]$, if the $i$th node of
$\ns{t}$ is some node $m$ then the $i$th node of $t'$ is of the form
$(m,q_j)$ for some file-state $q_j$, and

(c) the last node of $t'$ is of
the form $(m,q)$, for some $m$. \hfill$\Box$\\

Intuitively, the theorem above states that for all execution traces
that are due to record sequences that are accepted by the given input
automaton, control-flow paths taken by these traces are present in the
PFSG.

In the specific scenario where the PFSG is used to perform a dataflow
analysis, then the theorem above can be instantiated as follows.\\

\emph{Corollary:} Let $D$ be any \emph{sound} dataflow analysis
framework~\cite{Cousot77}, based on a semi-join lattice.  Let $d_0$ be a
given dataflow fact at program entry ($d_0$ is an element of $D$'s
lattice). For any node $n$ of the original CFG $G$, let $s(n)$ denote the
final fix-point solution at $n$ computed using $D$ on the CFG using initial
value $d_0$. Consider a PFSG for $G$ obtained using a given input automaton
$S$.  For any node $(n,q_i)$ of the PFSG, let $s(n,q_i)$ denote the final
fix-point solution at $n$ computed by $D$, but applied on the PFSG, using
the same initial value $d_0$.

\noindent Let
$s'(n) \equiv \bigsqcup_{\,q_i \mathrm{\ is\ a\ file-state\ of\ } S}\{s(n,q_i)\}$.

(a) $s'(n) \below s(n)$. [\emph{Precision}]

(b) $\gamma(s'(n))$ over-approximates the set of concrete
states that can arise at node $n$ when the program is run on input files
that conform to $S$. [\emph{Soundness}]\hfill$\Box$\\

\subsubsection{Precision ordering among PFSGs.}
As was clear from the discussion in this section, the PFSG produced by our
approach for a given CFG $G$ and input automaton $S$ is not fixed, but
depends on the selected underlying abstract interpretation $U$. The theorem
given above states that no matter what abstract interpretation is used as
$U$, the PFSG is sound (i.e., does not elide any paths that can be executed
due to record sequences that are accepted by $S$) as long as $U$ is
sound. However, the precision of the PFSG depends the precision of $U$. 

Given a CFG $G$ and an input automaton $S$, we can define a precision
ordering on the set of PFSGs for $G$ and $S$ that can be obtained using
different (sound) underlying domains $U_1, U_2$, etc.. A PFSG $P_1$ can be
said to be \emph{at-least as precise} as another PFSG $P_2$ if every edge
$(m,q_i) \rightarrow (n,q_j)$ in $P_1$ is also present in $P_2$. (Note that
this implies that every path in $P_1$ is also present in $P_2$.)\\

\emph{Theorem:} If an underlying domain $U_2$ is a consistent
abstraction~\cite{Cousot77} of another underlying domain $U_1$, then the
PFSG obtained for $G$ and $S$ using $U_1$ is at least as precise as the
PFSG obtained for $G$ and $S$ using $U_2$.\hfill$\Box$


\section{Implementation and evaluation}
\label{sec:Implementation}
\begin{figure}
  \centering
  \scriptsize
  \setlength{\tabcolsep}{5pt}
\begin{tabular}{|c|c|c|c|c|c|c|}
\hline
   &    & No. of    & \multicolumn{2}{c|}{Well-formed}   & \multicolumn{2}{c|}{Full}\\
 Prog.   & LoC     &  CFG      & \multicolumn{2}{c|}{Automaton}  & \multicolumn{2}{c|}{Automaton}\\\cline{4-7}
 name       &         & Nodes     & $|Q|$ & $|\Delta|$ & $|Q|$ & $|\Delta|$ \\
\hline
ACCTRAN   & 155     &  73    & 4   &  8 & 5    & 15  \\
\hline
SEQ2000	  & 219 	&  115  &  5  & 15 & 6    & 25   \\
\hline
DTAP      & 632 	&  275  & 5   & 6  & 10   & 41    \\
\hline
CLIEOPP   & 1421    &  900  & 21 & 48  & - & - \\
\hline
PROG1	  & 1177    &  762  & 8   & 16  & 12   & 47   \\
\hline
PROG2     & 1052    &  724  & 6    & 11  & 12   & 49     \\
\hline
PROG3     & 2780    &  1178  & 17  & 28  & 20   & 50   \\
\hline
 PROG4     & 49846   &  32258 & 13 & 34 & -  & -\\
 \hline
\end{tabular}
\caption{Benchmark program details}
\label{tab:pgmtab}
\end{figure}

We have targeted our implementation at Cobol batch programs. These are very prevalent in large enterprises~\cite{CompWorld2012}, and are based on a variety of standard as well as proprietary file formats. Another motivating factor for this choice is that one of the authors of this paper has extensive professional experience with developing and maintaining Cobol batch applications. We have implemented our analysis using a proprietary program analysis framework \emph{Prism}~\cite{PRISMRef}. Our implementation is in Java. We use the \emph{call strings} approach~\cite{sharirPneuliInterproc} for precise context-sensitive inter-procedural analysis. Cobol programs do not use recursion; therefore, we place no apriori bound on call-string lengths. 

Our analysis code primarily consists of an implementation of our generic analysis framework, as described in Section~\ref{ssec:approach:abstr-interpr-lift}.
We have currently not implemented our extension for data integrity
constraints that was described in Section~\ref{ssec:approach:persist}, nor have we implemented our PFSG construction approach (Section~\ref{sec:pfsg}).
We also have some lightweight scripts that process the fix-point solution emitted by the analysis to compute results for the specialization problem as well as the file conformance problem (see Section~\ref{sec:Applications}).

We ran our tool on a laptop with an Intel i7 2.8 GHz CPU with 4 GB RAM. 

\subsection{Benchmark programs}
\label{ssec:benchmark-programs}
We have used a set of eight programs as benchmarks for evaluation. Figure~\ref{tab:pgmtab} lists key statistics about these programs. The second and third columns give the sizes of these programs,  in terms of lines of code (including variable declarations) and in terms of number of (executable) nodes in the CFG (as constructed by Prism). The program ACCTRAN is a toy program that was used as a running example  in a previous paper~\cite{procinferenceWCRE2007}.  SEQ2000 is an example inventory management program used in a textbook~\cite{MurachsMainframeCOBOL} to showcase a typical sequential file processing program. The program DTAP has been developed by the authors of this paper. It is a payments validation program. The file-format it uses and the validation rules it implements are both taken from a widely used standard specification~\cite{DTASPECS}. The program CLIEOPP is a payment validation and transformation program. It was developed by a professional developer at a large IT consulting services company  for training purposes. The format and the validation rules it uses are from another standard specification~\cite{CLIEOP03}. PROG1 and PROG2 are real-world programs used in a bank for validating and reporting ``return'' payments sent from  branches of the bank to the head-office.  PROG3 and PROG4 are real-world programs from major multinational financial services companies. The program PROG3 is a \emph{format translator}, which translates various kinds of input records to corresponding output records. PROG4 reads data from a sequential master file, collects the data required for computing monthly interest and fee for each account, and writes this data out to various output files. The file formats used in these four real-world programs are proprietary.

Columns~4 and~5 in Figure~\ref{tab:pgmtab} give statistics about the well-formed automaton for each program.  For the programs ACCTRAN and SEQ2000 the respective original sources of these programs also give the expected input file formats.
For the real programs PROG1, PROG2, and PROG3, we derived the record types as well as well-formed automatons by going through the programs and guessing the intended formats of the input files to these programs. For program PROG4, the maintainers have provided us the file format specification.  In the case of programs DTAP and CLIEOPP, we constructed the record types as well as well-formed automatons from  their respective standard input-file specifications. In all cases we employed a precision-enhancing thumb-rule while creating the automatons, namely, that all incoming transitions into a file state be labeled  with the same type.

For most of the  programs we also constructed a \emph{full automaton}, to use in the context of ``over acceptance'' analysis. We created each full automaton using the corresponding well-formed automaton as a basis, following the basic procedure described in Section~\ref{ssec:appl:conformance}. Statistics about these full automatons are presented in the last two columns in Figure~\ref{tab:pgmtab}. We did not create a full automaton for CLIEOPP and PROG4, because the full automatons for these program turn out to be large and unwieldy to specify. Instead, we used the well-formed automatons in place of the full automatons in over-acceptance analysis, which can cause potential unsoundness.

We evaluate our approach in three different contexts -- its effectiveness in detecting file-format conformance violations in programs, its usefulness in specializing file-processing programs, and its ability to improve the precision of a standard dataflow analysis.

\subsection{File format conformance checking}
\label{ssec:file-form-conf}
\begin{figure}
  \centering
  \scriptsize
\begin{tabular}{|c|r|r|}
\hline
 Prog.  & \multicolumn{2}{|c|}{File format conformance warnings} \\\cline{2-3}
     Name   &  Under acceptance           & Over acceptance  \\
\hline
ACCTRAN   & 0 & 1 \\
\hline
SEQ2000	  & 3 & 1 \\
\hline
DTAP      & 0 & 1 \\
\hline
CLIEOPP   & 13 & * 0 \\
\hline
PROG1	  & 5 & 9 \\
\hline
PROG2    & 6 & 10 \\
\hline
PROG3  & 0 &  1  \\
 \hline
 PROG4  & 0 &  * 10  \\
 \hline
\end{tabular}
\caption{Conformance checking results}
\label{tab:FileConformanceCheck}
\end{figure}

As a first step in this experiment, we manually identified the \emph{rejection points} for each program. This was actually a non-trivial task, because each program had its own idioms for rejecting files. Some programs wrote warnings messages into log files, others used system routines for terminating the program, while others used Cobol keywords such as \pgm{GOBACK} and \pgm{STOP RUN}. Furthermore, since not every instance of a warning output or termination is necessarily due to file format related issues, we had to exercise care in selecting the instances that were due to these issues. We also manually added summary functions in our analysis for calls to
certain system routines that terminate the program: our summary functions treat these calls as returning a $\bot$ dataflow value for all file states, thus 
simulating termination. In this experiment we use CP (Constant Propagation) as the underlying analysis $U$ for our lifted approach.

Figure~\ref{tab:FileConformanceCheck} summarizes the results of this analysis. For each program, the second column captures the number instances of a file state of the well-formed automaton having a non-$\bot$ value at a rejection point. These are basically the under-acceptance warnings. 
The third column depicts the number of file states of the full automaton (excluding the final states of the original well-formed automaton) that reach the final point of the ``main'' procedure with a non-$\bot$ value.  These are basically the over-acceptance warnings.

The running time of the analysis was a few seconds or less on all programs except PROG4. On this very large program the analysis took 3700 seconds. 

\paragraph{Discussion of under-acceptance results.} A noteworthy aspect of these results is that four of the eight programs, namely, ACCTRAN, DTAP, PROG3, and PROG4 have been \emph{verified} as having no under-acceptance errors.
In the case of CLIEOP, some of the under-acceptance warnings turned out to be true positives during manual examination, in that the code contained programming errors that cause rejection of well-formed files.

We also manually examined one other program for which there were warnings -- SEQ2000. Although this program is a textbook program, it follows a
a complex idiom. Certain fields in certain record types in the input file format for this program are supposed to contain values that appear as primary keys in a sorted persistent table  that is accessed by the program. However, the well-formed automaton that we created does not capture this constraint, and is hence over-approximated
\footnote{This program uses sequential lookup on the persistent table,  which is an idiom that our persistent-stores extension  (Section~\ref{ssec:approach:persist}) does not support.}.
This caused false under-acceptance warnings to be reported.

\paragraph{Discussion of over-acceptance results.}
As is clear from the table, our implementation reports over-acceptance warnings on all the programs. (The numbers marked with a ``*'' are potentially lower than they should really be, because, as mentioned in
Section~\ref{ssec:benchmark-programs}, we did not actually use a full automaton for these two programs.) We manually examined four of these programs, and report our findings below.

Warnings reported for two of the programs -- DTAP and PROG4 -- turned out genuine. 
The input file-format for DTAP is similar to the one shown in Figure~\ref{fig:RunningOther}(a) (the difference is that it uses  single state $q_h$ in place of $\qsh$ and $\qdh$). This program happens to accept files that contain batches in which a header record and a trailer record occur back-to-back without any intervening item records, which is a violation of the specification. In the case of PROG4, when we discussed the warnings with the maintainers of the program, they agreed that some of them were genuine. However, at present, there is another program that runs before PROG4 in their standard workflow that ensures that ill-formed files are not supplied to PROG4.

In the case of  SEQ2000 and PROG3, the well-formed automatons were over-approximated. There is one other challenging idiom in SEQ2000, which also contributes to imprecision. Some of the routines that emit warnings emit file-conformance warnings when called from certain call-sites, and other kinds of warnings when called from other call-sites. Since we currently do not have a context-sensitive scheme to mark rejection points, we left these routines unmarked as a conservative gesture.

\subsection{Program specialization}
\begin{figure}
\centering
\scriptsize
\begin{tabular}{|c|c|c|c|c|}
\hline
  &   &  & Criterion- &  \\
 S. No      & Program     & Criterion      &  specific & Common   \\
       &    name      &  name         &  nodes    &  nodes       \\
\hline
1 & ACCTRAN	 & Deposit 	    &1   &   \\ \cline{1-4}
2 & ACCTRAN	 & Withdraw 	&30  & 41   \\
\hline
3 & SEQ2000	 & Add 	        &14  &    \\ \cline{1-4}
4 & SEQ2000	 & Change 	    &14  & 84 \\ \cline{1-4}
5 & SEQ2000	 & Delete 	    & 6  &    \\
\hline
6 & DTAP     & DDBank       &5   &  \\ \cline{1-4}
7 & DTAP     & DDCust       &7   & 216    \\
\hline
8 & DTAP     & CTBank  	    &5   &   \\\cline{1-4}
9 & DTAP     & CTCust  	    &7   & 216     \\
\hline
10 & CLIEOPP  & Payments 	&22  &  \\\cline{1-4}
11 & CLIEOPP  & DirectDebit	&98  & 622    \\
\hline
12 & PROG1    & Edit 	    &17  &   \\ \cline{1-4}
13 & PROG1    & Update 	    &215 & 511 \\
\hline
14 & PROG2    & Form        &3   & \\ \cline{1-4}
15 & PROG2    & Telex 	    &5   & 644\\ \cline{1-4}
16 & PROG2    & Modified    &5   & \\
\hline
17 & PROG3    & TranCopy    &37     &671     \\
\hline
12 & PROG4    & DAccts 	    &2  &   \\ \cline{1-4}
13 & PROG4    & MAccts 	    &1047 & 30825 \\
\hline
\end{tabular}
\caption{Specialization criteria and results}
\label{tab:DiffResults}
\end{figure}

The objective of this experiment is to evaluate the effectiveness of our approach in identifying program statements that are relevant to given criteria that are specified as specialization automatons.
In this experiment we used CP as the underlying analysis $U$.
We ran our tool multiple times on each program, each time with a different specialization criterion that we identified which represents a meaningful functionality from the end-user perspective. For instance, consider the program SEQ2000. The input file to this program consists of a sequence of \emph{request} records, with each request being either to
``Add'' an item to the inventory (which is stored in a persistent table), to ``Change'' the details of an item in the inventory, or to ``Delete'' an item from the inventory. A meaningful criterion for this program would be one that is concerned only about ``Add'' requests. Similarly, ``Change'' and ``Delete'' are meaningful criteria. Figure~\ref{tab:DiffResults} summarizes the results from this experiment. Each row in the figure corresponds to a program-criterion pair. The third column in the figure indicates the mnemonic name that we have given to each of our criteria. Note that for PROG3, the criterion TranCopy specializes the program
to process one of twelve kinds of input record types. While we have done the specialization with all 12 criteria, for brevity we report only one of them in the figure (i.e., TranCopy).

\paragraph{Results.}
For any criterion, the sum of the numbers in the fourth and fifth columns in the figure is the number of CFG nodes that were determined by our analysis as being relevant to the criterion (i.e., were reached with a non-$\bot$ value under some file state with the specialization automaton).  For instance, for ACCTRAN-Deposit, the number of relevant CFG nodes is 42 (out of a total of 73 nodes in the program -- see Figure~\ref{tab:pgmtab}). The fifth column indicates the number of (common) nodes that were relevant to \emph{all} of the criteria supplied, while the fourth column indicates the number of nodes that were relevant to the corresponding individual criterion but are \emph{not} common to all the criteria.
Note that in the case of DTAP we show commonality not across all four criteria, but within two subgroups each of which contains two (related) criteria. Also,
in the case of PROG3 the common nodes depicted are across all twelve criteria.

It is notable that in most of the programs the commonality among the statements that are relevant to the different criteria is high, while statements that are specific to individual criteria are fewer in number. Our belief is that in a program comprehension setting the ability of a developer to separately view common code and criterion-specific code  would let them  appreciate in a better way the processing logic that underlies each of these  criteria.

\paragraph{Manual examination.}
We manually examined the output of the tool to determine its precision. We did this for all programs except PROG1, PROG2, and PROG4 which had difficult control-flow as well as logic which made manual evaluation difficult. To our surprise, the tool was 100\% precise on \emph{every} criterion for four of the remaining programs -- ACCTRAN, CLIEOP, DTAP, and PROG3.
That is, it did not fail to mark as unreachable any CFG node that was actually unreachable (as per our human judgment) during executions on input files that conformed to the given specialization automaton. This is basically evidence that specialization automatons (in conjunction with CP as the underlying analysis) are a sufficiently precise mechanism to specialize file-processing programs.

The remaining one program is SEQ2000, for which, as discussed in Section~\ref{ssec:file-form-conf}, we have an over-approximated input automaton. Although the specialized program does contain extra statements that should ideally be removed, the result actually turns out to be 100\% precise relative to the given automaton.

\subsection{Precision improvement of existing analyses}
\label{ssec:results:prec-impr-exist}

As discussed in Section~\ref{ssec:analysis-issues}, there are scenarios
where one is interested in performing standard analyses on a program, but
restricted to paths that can be taken during runs on well-formed files
only. To evaluate this scenario we implemented two analyses. One is a
\emph{possibly uninitialized variables} analysis, whose abstract domain we
call \emph{Uninit}, wherein one wishes to locate references to variables
that have either not been initialized, or have been initialized using
computations that in turn refer to possibly uninitialized variables. The
second is a reaching definitions analysis, whose abstract domain we call
\emph{RD}. We ran each of these two analyses in two modes: a ``direct''
mode, where the analysis is run as-is, and a ``lifted'' mode, where the
analysis is done by lifting it with a well-formed automaton. In the lifted
mode, for \emph{Uninit} we used CP $\times$ \emph{Uninit} as the underlying
analysis $U$, while for \emph{RD} we used CP $\times$ \emph{RD} as the underlying analysis. (The CP component is required to enable path-sensitivity, as was illustrated in Figure~\ref{fig:RunningSoln}.) 

In the interest of space we summarize the results. With \emph{Uninit}, 82.5\% of all variable references in SEQ2000 were labeled as uninitialized in the direct mode, whereas only 25.7\% were labeled so in the lifted mode. For DTAP, the analogous numbers are 61.4\% and 9.6\%. In other programs the lifted mode performed only marginally better than the direct mode.

In the case of \emph{RD}, the total number of def-use edges computed by the lifted mode were 12\% below those computed by the direct mode for DTAP, 23\% below for CLIEOPP, and 15\% below for PROG2. In the other programs the reduction was marginal.

We do not have numbers for these experiments on the large program PROG4, as the cross-product domains (mentioned above) do not yet scale to programs of these sizes. On the other programs the direct analyses took anywhere from a few hundreds of a second to up to 18 seconds, while the lifted analyses took anywhere from a few tenths of a second to 180 seconds. 

We did a limited study of some programs where the lifted mode did not give significant benefit. Some of the causes of imprecision that we observed were array references, and calls to external programs, both of which we handle only conservatively. These confounding factors in these programs could not be offset by the precision improvement afforded by the input automatons.

\subsection{Discussion}
In summary, we are very encouraged by our experimental results. Except the two smaller programs -- ACCTRAN and SEQ2000 -- our benchmark programs are either real, or work on real formats and implement real specifications.

File-format conformance checking and program specialization are two novel problems in whose context we have evaluated our tool. The tool verified four programs as not rejecting any well-formed files, and found genuine file-format related errors in several other programs. The tool was very precise in the program specialization context.
Finally, it enabled non-trivial improvement in precision in the context of uninitialized variables or reaching definitions analysis on four of the eight programs.

\section{Related Work}
\label{sec:Related}
We discuss related work broadly in several categories.

\emph{Analysis of record- and file-processing programs}.  
There exists a body of literature, of which the work of Godefroid et al.~\cite{godefroid2008grammar}  and Saxena et al.~\cite{saxena2009loop} are representatives, on testing of programs whose inputs are described by grammars or regular expressions, 
via \emph{concolic execution}.  Their approaches are more suited for bug detection (with high precision), while our approach is aimed at conservative verification, as well as program understanding and transformation tasks. 

Various approaches have been proposed in the literature to recover record
types and file types from programs by program
analysis~\cite{komondoor_recovering_2007,caballero2007polyglot,cuiTupni08,repsConform11,kanadeConform12}. These approaches complement ours, by being potentially able to infer input automatons from programs in situations where pre-specified file formats are not available.

A report by Auguston~\cite{Auguston99decidabilityof} shows the decidability
of verifying certain kinds of  assertions in file-processing programs.

\emph{Program specialization.} Blazy et al.~\cite{SFAC:Blazy} describe an
approach to specialize Fortran programs using constant propagation.
There is a significant body of literature on
the technique of \emph{partial evaluation}~\cite{jones_book}, which is a
sophisticated form of program specialization, involving loop unrolling to
arbitrary depths, simplification of expressions, etc.
These approaches typically support only criteria on fixed sized program inputs.
Launchbury et al.~\cite{launchbury1991project} extend partial evaluation to
allow criteria on data structures. 
Consel et al.~\cite{consel1993parameterized} provide
an interesting variant of partial evaluation, wherein they propose
an abstract-interpretation based framework to specialize
functional programs with \emph{abstract values} such as signs, types and ranges. Our approach could potentially be framed  as an instantiation of their approach, with an input-automaton-based ``lifted'' lattice, and corresponding lifted transfer functions.

\emph{Program slicing}.
Program slicing~\cite{xuSlicingSurvey2005} is widely applicable in software
engineering tasks, usually to locate the portion of a program that is
relevant to a criterion. The \emph{constrained} variants of program
slicing~\cite{field1995parametric,Canfora1998,HarmanPrePost2001}
provide good precision in general, at the cost of being potentially
expensive. Existing  approaches for constrained slicing
do not specifically support
constraints on the record sequences that may appear in input files of
file-processing programs. Our ``lifted'' reaching-definitions analysis, which we described in Section~\ref{ssec:results:prec-impr-exist}, enables this sort of slicing.

\emph{Typestates.} There is a rich body of literature in specifying and
using \emph{type states}, with the seminal work being that of Strom et
al.~\cite{strom1986typestate}. In the context of analyzing file-processing programs, type-state automatons have been
used to capture the state of a file (e.g., ``open'', ``closed'', ``error'')~\cite{manuvirEsp2002,fischer2005joining}. To our knowledge ours is the first work in this space to use automatons to encode properties of the prefix of records read from a file.

\emph{Shape analysis.} Shape analysis~\cite{Sagiv} is a precise but
heavy-weight technique for verifying shapes and other properties of
in-memory data structures. While at a high level a data file is similar
to an in-memory list, the operations used to traverse files and in-memory
data structures are very different. To our knowledge shape analysis has not
been used in the literature to model the contents and states of files as
they are being read in programs. It would be an interesting topic
of future work to explore in-depth the feasibility of such an approach.

\section{Conclusions and Future Work}
\label{sec:concl-future-work}
We presented in this paper a novel approach to apply any given abstract
interpretation on a file-processing program that has an associated input
file-format. The file-format basically enables our approach to elide
certain paths in the program that are infeasible as per the file format,
and hence enhance the precision and usefulness of the underlying
analysis. We have demonstrated the value of our approach using
experiments, especially in the context of two novel applications: file
format conformance checking, and program specialization.

A key item of future work is to allow richer constraints on the data in the
input file and persistent tables; for instance, general logical
constraints, constraints expressing sortedness, etc., would be useful in
many settings to obtain enhanced precision and usefulness. Also, we would
like to investigate our techniques on domains other than batch programs;
e.g., to image-processing programs, XML-processing programs, and web-based
applications.

\appendix
\section{Appendix}

\subsection{Precision of our approach}
\label{ssec:approach:precision}
We discuss here  the precision of our approach, which was alluded to in Section~\ref{ssec:approach:soundn-prec-compl}, in more detail.  An  $L$-\emph{solution} is a function from program points in the given program $P$ to dataflow values from the lattice $L$.  A $(Q,L)$-solution  is a function from program points to functions in the domain $Q \rightarrow L$, where $Q$ is the set of file states in an input automaton.

Given a $(Q,L)$ solution $g$ and an $L$-solution $f$ (for the same program $P$), we say that $g$ is \emph{more precise} than $f$ iff at each program point $p$:
  $$\sqcup_{q \in Q}\{(g(p))(q)\} \sqsubseteq_L f(p)$$

Note that we are actually using ``more precise'' as shorthand for ``equally precise or more precise''.

Let $f_1$ be the fix-point $L$-solution obtained for program $P$ by directly using the underlying analysis $U$. Our first key result is as follows.

\begin{theorem}\label{thm:autom-more-precise}
If $S$ is \emph{any} input automaton for $P$ with set of files states $Q$, then the $(Q,L)$ fix-point solution computed by our approach using $S$ and using the underlying analysis $U$ is more precise than fix-point solution computed by $U$ directly.
\end{theorem}

Intuitively, the above theorem  captures the fact that the path-sensitivity that results from tracking different dataflow values (from lattice $L$) for different file states causes increase in precision. Note that the theorem above does not touch upon soundness. In order to ensure soundness, $S$ would additionally need to accept all well-formed files or all files, depending on the notion of soundness that is sought.

A different question that naturally arises is, if there are multiple candidate well-formed automatons that all accept the same set of well-formed files, will they all give equally precise results when used as part of our analysis? The answer, in general, turns out to be ``no''.
It can also be shown that if an input automaton accepts a smaller set of files than another automaton, then the first automaton need not necessarily give more precise results than the second one on all programs.  In fact, precision is linked both to the set of files accepted as well as to the \emph{structure} of the automatons themselves.

In order to formalize the above intuition, we first define formally the notion of a precision ordering on different solutions for a program $P$ using different input automatons. A $(Q_1,L)$ solution $g_1$ is said to be more precise than another $(Q_2,L)$ solution $g_2$ iff at each program point $p$, for each file state $q_1 \in Q_1$, there exists a file state $q_2 \in Q_2$ such that:
$$(g_1(p))(q_1) \sqsubseteq_L (g_2(p))(q_2)$$

We then define a notion of \emph{refinement} among input automatons for the same program $P$.  We say that automaton $S_2$ = $(Q_2,
\Sigma_2, \Delta_2, q_{s2}, Q_{e2})$  is a \emph{refinement} of automaton $S_1$ = $(Q_1, \Sigma_1, \Delta_1,
q_{s1}, Q_{e1})$ iff there exists a mapping function $m: Q_2 \rightarrow Q_1$ such that:\\

\noindent- $m(q_{s2})$ = $q_{s1}$, and

\noindent- For each transition $p_2 \rightarrow q_2$ in $\Delta_2$ labeled with some symbol $s_2 \in \Sigma_2$: There exists one or more transitions from  $m(p_2)$ to $m(q_2)$ in     $S_1$. Furthermore, if $s_1$ is the label on any  of these transitions,  then either (1) $s_2$ and $s_1$ are both $\eof$, or (2) $s_2$ and $s_1$ are both types, and $s_2$'s constraint implies $s_1$'s constraint. \\

If an input automaton $S_2$ is a refinement of an input automaton $S_1$, then the following two properties can be shown to hold: (1) each accepting state of $S_2$ is mapped by $m$ to an accepting state of $S_1$, and (2) the set of files accepted by $S_2$ is a subset of files accepted by $S_1$. Now, our main result on precision ordering of input automatons is as follows.

\begin{theorem}\label{thm:refinement}
For any program $P$ and for any given underlying analysis $U$, if an input automaton $S_2$ for $P$ is a refinement of an input automaton $S_1$ for $P$, then the fix-point solution computed by our approach using $S_2$ and $U$ is more precise than the fix-point solution computed by our approach using $S_1$ and $U$.
\end{theorem}

An important take away from the above theorem is: When there is a choice between  two input automatons that accept the same set of files (e.g., two different well-formed automatons accepting the same set of well-formed files), if one of them is a refinement of the other then the refined automaton will give more precision than one of which it is a refinement. 

\subsection{Checking over-acceptance errors}

We discuss here a procedure to extend a given well-formed automaton $S$ = $(Q, \Sigma = \mathcal{T} \cup \{\eof\}, \Delta, q_s, Q_e)$ into a full automaton.  We first create a new type named ``NA'' (none of the above), and associate with it a constraint that lets it cover all records that are not covered by any of the types in the original set of types $\mathcal{T}$. We also add a new file state to the well-formed automaton, which
we denote as $q_y$ in this discussion. Let $\mathcal{T'} \equiv \mathcal{T} \cup \{\mathrm{NA}\}$, and $Q' \equiv Q \cup \{q_y\}$. For every state $q$ in $Q'$, and for every type $t$ in $\mathcal{T'}$, if there is no transition labeled $t$ out of $q$ we add a transition from $q$ to $q_y$ labeled $t$. Finally, we add one more new file state $q_x$ to the automaton, make it a final state, add $\eof$ transitions from all non-final states to this state. The intuition behind this construction is that $q_x$ accepts all ill-formed files, while $q_y$ accepts all record sequences that are not prefixes of well-formed files. 

\end{document}